\documentclass[letterpaper]{article}
\usepackage[T1]{fontenc}

\usepackage{geometry}
\usepackage{setspace}

\usepackage{graphicx}
\usepackage{float}
\newfloat{scheme}{htbp}{los}
\floatname{scheme}{Scheme}
\newfloat{graph}{htbp}{loh}

\usepackage{bm}% bold math
\usepackage{lineno}

\usepackage{chemformula} % Formulas using \ch{}
\usepackage[version = 4]{mhchem} % Formulas using \ce{}

\newcommand{\beq}{\begin{equation}}
\newcommand{\eeq}{\end{equation}}
\newcommand{\beqa}{\begin{eqnarray}}
\newcommand{\eeqa}{\end{eqnarray}}
\newcommand{\kB}{\mbox{$k_{\rm B}$}}
\newcommand{\kBT}{\mbox{$k_{\rm B}T$}}
\newcommand{\TR}{\mbox{$T_{\rm R}$}}

\newcommand{\simless}{\stackrel{<}{\sim}}

\usepackage{authblk}
\author[1]{Noriko Akutsu}
\author[2]{Yoshihiro Kangawa}
\affil[1]{Graduate School of Energy Science, Kyoto University, 
Yoshida-honmachi, Sakyo-ku, Kyoto 606-8501, Japan
}
\affil[2]{Research Institute for Applied Mechanics, Kyushu University,  Kasuga, Fukuoka, 816-8580, Japan}
\title{
Intrinsic Step Jamming in Nanometer-Scale KPZ-like Rough Surfaces under Interface-Limited Crystal Growth and Retreat
}
\date{*Email: akutsu.noriko.3y@kyoto-u.ac.jp}

\begin{document}

\maketitle

\begin{abstract}

We investigate intrinsic step jamming at the nanometer scale on Kardar--Parisi--Zhang (KPZ)-like, kinetically roughened two-dimensional crystal surfaces embedded in three-dimensional space, using Monte Carlo simulations of a restricted solid-on-solid (RSOS) model.
In this model, transport processes such as surface and volume diffusion, elastic interactions, the Ehrlich--Schwoebel effect, and explicit step--step interactions are excluded.
Intrinsic step jamming emerges under nonequilibrium driving as a consequence of asymmetric atomic attachment and detachment at step edges under the RSOS non-penetrability constraint, and persists on nanometer-scale surfaces, with step congestion becoming pronounced at characteristic length scales of approximately 1.6 nm for (111) stepped surfaces in the present model.
This inhomogeneity is directly reflected in the terrace width histogram and corresponds to transient nanoscale traffic-jam-like clusters rather than conventional step bunching.
The behavior is analogous to jamming in the asymmetric simple exclusion process (ASEP) at the phenomenological level, including multi-lane variants.
Intrinsic step jamming governs surface morphology, producing bell- or cup-shaped surface undulations associated with positive or negative skewness in the surface-height-difference distribution.
Strategies to suppress intrinsic step jamming are also discussed.

\end{abstract}

%\section*{Keywords}

%Kinetic roughening, Monte Carlo method, Kardar--Parisi--Zhang universal class, two-dimensional nucleation, asymmetric simple exclusion process

%\section*{Abbreviations}

%Some journals require a list of abbreviations: these normally should be given
%immediately after the keyswords (if required).


%\tableofcontents
%\linenumbers

\section*{Introduction}

%background
In crystal growth theory, surface steps are traditionally regarded as the key microscopic degrees of freedom on smooth crystal surfaces, where individual steps are well defined, and their motion directly determines surface evolution. 
Under nonequilibrium conditions, such step dynamics govern growth modes and morphological stability as long as the surface remains smooth \cite{bcf, frank58, chernov61, pimpinelli, saito, sato2000, misbah10, einstein15book, akutsu15book, Kotur21}. 
However, this conventional picture implicitly assumes a clear separation between step-dominated smooth surfaces and kinetically roughened surfaces described by continuum height fluctuations.

% kinetic roughening
In statistical mechanics, kinetic roughening is commonly discussed in terms of continuum height fluctuations and universal scaling behavior \cite{barabasi,vicsek}, where a surface is assumed to obey the Family--Vicsek scaling function at non-equilibrium as $W(L,t)= L^{\alpha} f(t/L^z), \quad z=\alpha/\beta$, 
where $W(L,t)$ is the standard deviation of the surface height at time $t$ for a linear system size $L$, $f(x)$ is a scaling function.
The scaling function $f(x)$ converges to 1 for $x \rightarrow \infty$ and converges to $x^\beta $ for $x \rightarrow 0$, and $\alpha$, $\beta$, and $z$ are the roughness, growth, and dynamic exponents.
In crystal growth, kinetic roughening often refers to surfaces on which individual steps are no longer well defined. \cite{bcf,van veenendaal,cuppen}. 
The surface is considered to evolve in adhesive growth, in which the growth rate increases linearly with the driving force. 
In this manner, the same term may refer to different physical situations depending on the context.

%%%%% FIGURE  %%%%%

\begin{figure*}%[htbp]
%\begin{center}
\centering
\includegraphics[width=15cm,clip]{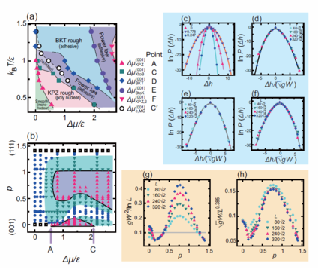}%
%\end{center}
\caption{
Representative kinetic roughening behaviors under interface-limited steady crystal growth and retreat.
(a) Kinetic roughening diagram for the (001) surface ($p=0$) as a function of temperature and driving force, showing smooth, Berezinskii-Kosterlitz-Thouless (BKT) rough, and Kardar-Parisi-Zhang (KPZ)-like kinetically rough regimes.
(b) Kinetic roughening diagram for inclined surfaces, illustrating the dependence of kinetic roughening behavior on surface slope ($\kBT/\epsilon = 0.4$).
For the diagrams at several other temperatures, please refer to Ref. \cite{akutsu25-2}.
(c)-(f) Logarithmic plots of surface-height-difference distributions (SHD) for selected conditions. $\kBT/\epsilon =0.4$. $\Delta \mu/\epsilon = 1.4$.
(g),(h) Scaled surface widths as functions of system size used to classify kinetic roughening regimes.
These panels provide background context for the present study and define the kinetic roughening landscape in which the microscopic behavior of surface steps is investigated. Panels (a) to (h) are reproduced or adapted from Refs. \cite{akutsu25-2, akutsu23, akutsu24-2} under the Creative Commons Attribution 4.0 International License (CC BY 4.0).
\label{kr-diagram}
}
\end{figure*}
%%%%% FIGURE  %%%%%

%Definitions of rough and smooth surfaces

In this paper, we distinguish two different notions of surface roughness that are essential for discussing step dynamics. 
Thermodynamic roughness \cite{chui76, knops, beijeren77, weeks80, jayaprakash83, akutsu87} is defined on the basis of the criteria at equilibrium
\beqa
&&W\rightarrow \infty \quad \text{as} \quad L \rightarrow \infty \quad (\text{rough}), %\qquad
\nonumber \\
&&W\rightarrow \text{const.} \quad \text{as} \quad L \rightarrow \infty \quad (\text{smooth}),
\label{eq:rough_smooth}
\eeqa
where $W$ is the surface width and $L$ is the linear system size. 
In contrast, we define a surface as atomically smooth when individual surface steps are well identified, whereas a surface is atomically rough \cite{nishinaga89} when step edges are no longer distinguishable at the microscopic scale. 
Atomically rough surfaces are associated with adhesive growth.
These definitions are introduced to avoid ambiguity and are used consistently throughout this work.

%KPZ kinetic roughening
Previous studies have shown that, with steady growth under interface-limited conditions, kinetically roughened surfaces exhibit Kardar--Parisi--Zhang (KPZ)-like roughness with the exponent $\alpha = \alpha_{\rm KPZ} = 0.3869$ \cite{kpz, pagnani15, takeuchi18}, depending on surface orientation, driving force, and temperature (Fig.~\ref{kr-diagram} (a) and (b), the kinetic roughening diagrams, with $\kBT/\epsilon = 0.4$ for (b)) \cite{akutsu23, akutsu24-2, akutsu25-2}.
Here, the interface-limited condition means that atoms in the environmental phase reach the surface rapidly enough. 
So far, KPZ rough surfaces are generally expected at temperatures higher than the thermal roughening transition temperature and during growth far from equilibrium because the roughness exponent $\alpha$ of the thermal roughening transition equals zero, and the self-affine KPZ rough surface was considered to occur near the self-similar dendritic growth where $\alpha=1$.
In contrast to expectations, KPZ-like roughened surfaces form at low temperatures and near equilibrium, when the surface is atomically smooth yet thermodynamically rough.

% alpha =0
Here, we consider the two cases for the roughness exponent $\alpha=0$: one is $W^2 \propto \ln L$, and the other is $W \propto \ln L$, which are regarded as Berezinskii--Kosterlitz--Thouless (BKT)\cite{ chui76, knops, beijeren77, weeks80,jayaprakash83, akutsu87, berezinskii71, kt}  rough and Edwards--Wilkinson (EW) rough \cite{barabasi}, respectively.
Since the Monte Carlo data were better fitted to $W^2 \propto \ln L$ than $W \propto \ln L$, we regard the $\alpha =0$ region as BKT rough.
The gray line in Fig.~\ref{kr-diagram} (g) is $W^2 /\ln L = 1/\pi^2$, which is the BKT universal amplitude at the roughening transition temperature of the (001) surface $\TR^{(001)} $.
The line is often used as a criterion for whether the terrace surface is rough at equilibrium \cite{dennijs}. 
This BKT rough region connects to the expected BKT rough region at a higher temperature than $\TR^{(001)}$ (Fig.~\ref{kr-diagram} (a)).

%KPZ subclasses 
As seen from Fig.~\ref{kr-diagram} (a) and (b), the KPZ-like roughened areas are separated by the BKT rough area, which suggests there exist subclasses in the KPZ class for surface kinetic roughening phenomena.
A useful diagnostic quantity for characterizing such kinetically roughened states is the surface-height-difference distribution (SHD) $P(\Delta h)$ \cite{akutsu25-2, takeuchi10, takeuchi11, halpin, oliveira13, almeida14, almeida17} which is equivalent to the local surface slope distribution.
The respective KPZ-like areas separated by the BKT rough area have their own skewness and kurtosis values, respectively, and the KPZ-like areas are then judged to be subclasses of the KPZ class.
These areas are assigned as follows: KPZ-like1 for the area near $p=0$, KPZ-like2 for the area around $p=\sqrt{2}/2$, and KPZ-like3 for the area at high temperature and driving force. 
These subclasses are characterized by different step configurations and growth modes \cite{akutsu25-2} and inherit the one-dimensional interface subclasses \cite{takeuchi10, takeuchi11}.

%SHD
Here, the sign of the skewness of $P(\Delta h)$ provides a simple criterion to identify asymmetric surface undulations: positive and negative skewness correspond to bell-shaped and cup-shaped surface morphologies, respectively \cite{akutsu25-2}.
This classification plays a central role in the following sections, where it is used to diagnose the emergence and suppression of ``intrinsic step jamming'', defined later in this section.
In fact, the skewness and kurtosis of the SHD, which were defined and systematically analyzed in our previous work \cite{akutsu25-2}, are used here as diagnostic measures, and their numerical values are not repeated in the present study.

In our previous work \cite{akutsu25-2}, the terrace width histograms show that there is an unknown inhomogeneity arising in the step density on KPZ-like rough surfaces, somewhat similar to step bunching.
%step bunching in continuous model
Step bunching and step meandering have long been observed in crystal growth and etching processes and have often been described using continuum models \cite{chernov61, pimpinelli, saito, sato2000, misbah10}.
The continuum descriptions typically rely on diffusion fields \cite{chernov61, pimpinelli, saito}, electromigration \cite{stoyanov91, natori92, stoyanov98-1, pimpinelli02}, kinetic asymmetries such as the Ehrlich--Schwoebel (ES) effect \cite{ehrlich, schwoebel} to account for the emergence of step bunching \cite{sato2000, Kotur21}, and external fields \cite{garcia}.
The step bunching associated with diffusion phenomena occurs at length scales comparable to or larger than the diffusion length.
The above step-bunching phenomena disappear at equilibrium.
%macrostep formation
For step bunching to remain at equilibrium, effective step--step interactions \cite{einstein15book, akutsu15book, song94, akutsu16, akutsu19, akutsu22} or thermodynamical adsorption effects \cite{akutsu01-2, akutsu03, akutsu09-2} are required; in such cases, step bunching or macrostep formation is realized as two-surface coexistence at equilibrium due to an anomaly in the anisotropic surface tension (surface free energy per area).
%shenoy
For more microscopic-scale step bunching, a combination of short-range attractive and long-range repulsive interactions between steps has been theoretically shown to produce a regular train of bunched steps \cite{shenoy2000, minoda}. 
The bunched step height depends on the strength of the repulsive step--step interaction.

% difference

However, the mechanism causing the step density inhomogeneity in our previous work \cite{akutsu25-2} is fundamentally different from those of the above-mentioned step-bunching phenomena, because the known step-bunching mechanisms are absent from the present model. 
Hence, we refer to this phenomenon as intrinsic step jamming to distinguish it from diffusion-driven or elastic-interaction-induced step bunching.

%step jamming

In this work, we identify a previously unexplored intrinsic step jamming mechanism that emerges on kinetically roughened crystal surfaces under interface-limited growth and retreat, even in the absence of diffusion, elastic interactions, or explicit step--step forces.
The mechanism originates from biased atomic attachment and detachment at step edges combined with the non-penetrability constraint imposed by the solid-on-solid (SOS) geometry.
As a consequence, surface steps accumulate into transient high-density areas, that is, transient nanoscale step clusters, giving rise to congestion-like dynamics that are characteristic of jamming.
Such congestion bears close structural similarity to jamming phenomena in driven exclusion processes, most notably the asymmetric simple exclusion process (ASEP) \cite{blythe, ishiguro}.
Depending on the direction of the driving force for crystal growth or retreat, this intrinsic step jamming manifests as bell-shaped or cup-shaped surface undulations, respectively, leading to KPZ-like roughening through step dynamics.

The specific quantities characterizing the intrinsic step jamming will be summarized at the top of subsection ``Intrinsic step jamming''.
Three approaches to suppress intrinsic step jamming will also be discussed in the subsection ``Suppression of intrinsic step jamming''.

%%%%% FIGURE  %%%%%

\begin{figure}%[h]%[htbp]
%\begin{center}
\centering
\includegraphics[width=8.5cm,clip]{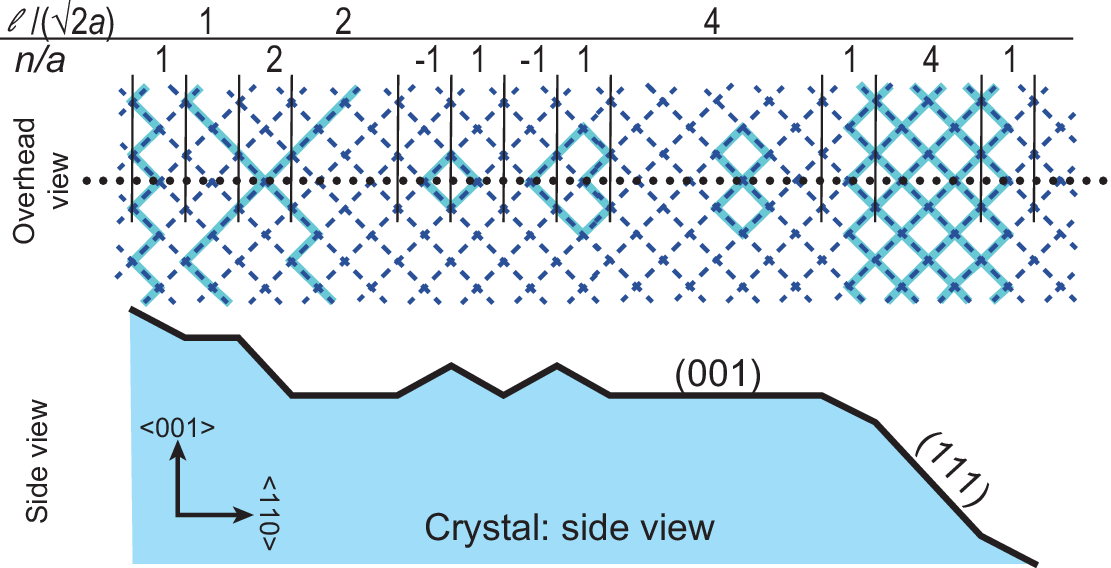}%
%\end{center}
\caption{Terrace width $\ell$ and step height $n$ with the (001) terrace.
Upper panel: an overhead view of a surface.
Broken lines show the lattice structure.
Thick, light-blue lines show surface steps.
Lower panel: a side view of the surface along the dotted line in the upper panel.
}  
\label{fig_deltahDef}
\end{figure}
%%%%% FIGURE  %%%%%

\section*{Microscopic model \label{sec_model}}

%the Hamiltonian
We used a restricted solid-on-solid (RSOS) model \cite{dennijs, yakutsu89} on a square lattice.
Here, ``restricted'' means that height differences between nearest-neighbor (nn) sites are limited to zero or $\pm$ one.
The surface energy near a (001) surface, including (001) terrace roughness, is expressed by the following discrete Hamiltonian:
\beq
%&&
{\cal H}_{\rm RSOS} = {\cal H}_{\rm conf} +{\cal H}_{\rm drive}, 
 \label{hamil}
\eeq
where ${\cal H}_{\rm conf} $ is the configuration part of the energy, and ${\cal H}_{\rm drive}$ is the chemical potential part driving crystal growth/retreat.
These terms are expressed explicitly as follows: 
\beqa
%&&
  {\cal H}_{\rm conf} &=&{\cal N}\epsilon_{\rm surf}+ \sum_{n,m} \epsilon 
[ |h(n+1,m)-h(n,m)|  \nonumber \\
&&
+|h(n,m+1)-h(n,m)|],   \nonumber \\
{\cal H}_{\rm drive} &=&- \sum_{n,m} \Delta \mu \ h(n,m),  
 \label{hamil1}
\eeqa
 where $h(n,m)$ is the surface height at site $(n,m)$, ${\mathcal N}$ is the total number of lattice points on the surface, $\epsilon_{\rm surf}$ is the surface energy per unit cell on the flat (001) surface, and $\epsilon$ is the microscopic ledge energy stabilizing the (001) surface.
The summation with respect to $(n,m)$ is taken over all sites on the square lattice.
$\Delta \mu$ is the driving force for crystal growth, defined by $\Delta \mu= \mu_{\rm ambient}- \mu_{\rm crystal}$, where $\mu_{\rm ambient}$ and $\mu_{\rm crystal}$ are the chemical potentials of the ambient phase and the crystal, respectively.
If the ambient phase is an ideal gas, $\Delta \mu= \kBT \ln P/P_{\rm eq}$ \cite{kangawa01, akutsu20, akutsu92}, where $\kB$ is the Boltzmann constant, $P$ is the gas pressure, and $P_{\rm eq}$ is the gas pressure at equilibrium.
%note for readers in the crystal growth field
Although $\epsilon_{\rm surf}$ and $\epsilon$ depend on temperature because they correspond to the surface free energy in a quantum mechanical atomic model \cite{kempisty19}, we assume they are constant throughout this work.

%%%%% FIGURE  %%%%%

\begin{figure}%[h]%[htbp]
%\begin{center}
\centering
\includegraphics[width=6cm,clip]{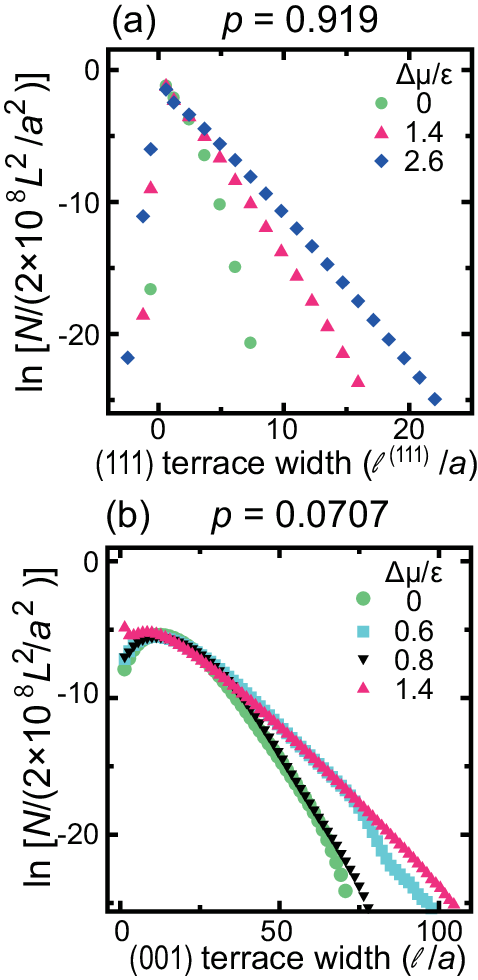}%
%\end{center}
\caption{Terrace width histograms (TWHs).
$\kBT/\epsilon = 0.4$. $L=320 \sqrt{2}a$.
(a)  Logarithmic histogram of the (111) terrace width.
$\ell^{(111)} /a= \sqrt{3/2} \ n $.
The mean terrace width $\ell_0^{(111)} = \sqrt{3/2} \ p/(\sqrt{2}- p) = 2.27 a$.
The relationship between $\ell^{(111)}$, $n$, and $p$ is derived in the sub-Sec. ``Terrace width for (111) steps'' in the Sec. ``Methods''.
 (b) Logarithmic histogram of the (001) terrace width.
 The mean terrace width $\ell_0 = a/p = 14.14 a$.
}  
\label{fig_terraceW}
\end{figure}
%%%%% FIGURE  %%%%%

% update in MC 
We studied surfaces using the Monte Carlo method for a non-conserved system, varying the number of crystal atoms.
The transition probability is 1 for $\Delta E= E_f-E_i \leq 0$ and $\exp (-\Delta E/\kBT)$ for $0<\Delta E$, where $E_i$ is the surface energy of the initial configuration, and $E_f$ is the surface energy of the proposed configuration.
The surface energy was calculated using Eqs. (\ref{hamil}) and (\ref{hamil1}).

%averaging method
The first $2 \times 10^8$ Monte Carlo steps per site (MCS/site) are discarded to study the steady state. 
Then, the quantities are averaged over the following $2 \times 10^8$ MCS/site.
%definitions
The number of steps $N_{\rm step}$ is fixed to give a surface slope of 
\beq
p=N_{\rm step}a/L,
\eeq
where $a=1$ is the lattice constant.
For inclined surfaces, a squared surface width, which is the variance of the surface height, is calculated as 
\beq
gW^2 =\langle \langle [h(\tilde{x}, \tilde{y}, t)- \langle h(\tilde{x}, t)\rangle_{\tilde{y}}]^2\rangle_{\tilde{y}} \rangle_{\tilde{x}}. 
\label{eq_w2def}
\eeq
Here, $W$ is the surface width normal to the inclined surface, defined as the standard deviation of the surface height, and $g= 1+p_x^2 +p_y^2$, with $p_x=\partial \langle h \rangle /\partial x$ and $p_y=\partial \langle h \rangle/\partial y$, is a geometrical factor, $\tilde{x}$ and $\tilde{y}$ are the $[ 110 ]$ and $[ \bar{1}10 ]$ directions (Fig.~\ref{fig_deltahDef}), respectively, and $\langle \cdot \rangle_{\tilde{y}}$ and $\langle \cdot \rangle_{\tilde{x}}$ are the averages over the $\tilde{y}$ and $\tilde{x}$ directions.

\section*{Results and Discussions}

\subsection*{Terrace width histogram (TWH) \label{sec_terraceW}}

%entropic repulsion
It is known that the terrace width distribution on stepped surfaces
is often described by the Wigner--surmise form derived from
random-matrix theory \cite{einstein99, giesen00, einstein}.
By fitting the terrace width distribution to the Wigner--surmise form, the strength of the Calogero--Sutherland-type repulsion \cite{calogero, sutherland}
($\propto \ell^{-2}$) can be measured as the parameter
$\tilde{A}_{\rm eff}$ in their terminology.
The Calogero--Sutherland-type step--step repulsion is considered to come from the elastic interaction \cite{alerhand} and the entropic repulsion \cite{pimpinelli02, einstein} in the continuous theory.
In the present surface energy Eqs. (\ref{hamil}) and (\ref{hamil1}), the repulsive term of the $\propto \ell^{-2}$ type is not included explicitly.
However, coarse-grained descriptions may still capture
effective entropic repulsion arising from step fluctuations.

%definition of terrace W
The (001) terrace width $\ell$ at $\tilde{y}$ is assigned as shown in Fig.~\ref{fig_deltahDef}.
The (111) terrace width $\ell^{(111)}$ is determined from the (001) step height $n$.
The microscopic orientation of the side surface of a step with a (001) terrace is (111) for $1<|n|$. 
 The terrace width histograms (TWHs) $N$ are calculated over $2 \times 10^8$ MCS/site, where $N$ is the frequency at terrace width $\ell$ or $\ell^{(111)}$.
 The logarithmic TWHs, divided by $2 \times 10^8 (L/a)^2$, are shown in Fig.~\ref{fig_terraceW}.  
Here, $\ell_0$ is defined as the mean terrace width $\ell_0=a/p$ for steps with (001) terraces.
For steps with (111) terraces, $\ell_0^{(111)}= \sqrt{3/2} \ n_0 a$ with $n_0= p/(\sqrt{2}-p)$ (see Sub-sec.~``Terrace width for (111) steps'' in Sec.~``Methods''). 
Hence, the mean terrace width $\ell_0^{(111)}$ for $p = 416/(320\sqrt{2}) \approx 0.919$ equals $2.27a$, while $\ell_0$ for $p= 1/(10 \sqrt{2}) \approx 0.0707=\tan \theta$, where $\theta$ is the off-angle from the (001) surface towards the (111) surface as $\theta = 4.04$ degrees,  equals $14.14a$.

%at equilibrium
At equilibrium ($\Delta \mu =0$), the TWHs for the (111) and (001) terrace widths  (Fig.~\ref{fig_terraceW}) have approximately a truncated Gaussian shape. 
It is noteworthy that the obtained TWHs deviate from the Wigner--surmise distributions, indicating that the effective entropic repulsion cannot be extracted from the present data.
Physically, step fluctuations dominate in the capillary wave, resulting in the truncated Gaussian distribution at equilibrium.
For small $\Delta \mu$, random zig-zag fluctuations associated with capillary waves exist around the mean terrace width $\ell_0$ or  $\ell_0^{(111)}$.

%%%%% FIGURE  %%%%%

\begin{figure*}%[t]%[htbp]
%\begin{center}
\centering
\includegraphics[width=13cm,clip]{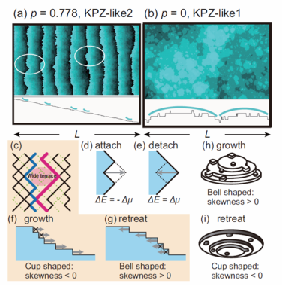}%
%\end{center}
\caption{
Example of surface undulations.
(a)  $\kBT/\epsilon = 0.4$, $L=80 \sqrt{2}a$.  $N_{\rm step} = 88$.
$\Delta \mu/\epsilon = 1.4$.  
In the white ellipses, typical surface undulations are visible.
(b)  $\kBT/\epsilon = 0.63$, $L=320 \sqrt{2}a$.
$\Delta \mu/\epsilon = 0.8$.  
(c) An example of a wide (001) terrace.
(d), (e) Attachment and Detachment of an atom at a configuration at a half-crystal site, respectively.
(f), (g) An example of a side view of the KPZ-like2 rough surface when it grows/recedes.
(h), (i) An example of a perspective view of the KPZ-like1 rough surface when it grows/recedes.
Detailed explanations are given in Sec. ``Intrinsic step jamming''.
}  
\label{surfdatCut}
\end{figure*}
%%%%% FIGURE  %%%%%

%characteristic of the (111) terrace W at equilibrium
For the (111) terrace width (Fig.~\ref{fig_terraceW} (a)), the frequencies of wide terraces increase as $\Delta \mu$ increases monotonically.
When the surface is in the KPZ-like2 region (Fig.~\ref{kr-diagram} (b)), the logarithm of the TWH for (111) terraces decreases linearly.
It can be clearly seen that the frequencies at the widths $4a \simless \ell^{(111)}$ deviate from the equilibrium ones.
Assuming a typical lattice constant of $a = 0.4$ nm, this clear increase occurs at lengths of 1.6 nm.
%Abstract
This value provides the basis for the characteristic
nanoscale length of about 1.6 nm mentioned in the Abstract.

%STD of TWH
In our previous work \cite{akutsu25-2}, we analyzed the TWH for the (111) terrace width by assuming the shape of the TWH to be the product of Gaussian and exponential decay functions:
\beq%a
N/N_{\rm max}= c_1 \exp[-(\ell^{(111)}/\ell_0^{(111)}-1)^2/\{2(\sigma/\ell_0^{(111)})^2\}% \nonumber \\
%&&
 - c_2 \ell^{(111)}/\ell_0^{(111)}] \label{eq_histo}
\eeq%a
where $N_{\rm max}$ is the maximum value of $N$.
Using the least square method, the Monte Carlo results are fitted to Eq. (\ref{eq_histo}) around $\ell^{(111)} \sim \ell_0^{(111)}$ by adjusting $\sigma$, $c_1$, and $c_2$.
Then we found that the stepped surface becomes KPZ-like2 when $\sigma$ exceeds $\ell_0^{(111)}$.
$\sigma$ in the KPZ-like2 rough region increases as $\Delta \mu$ increases. 
This means that the fluctuation width of the terrace width becomes comparable to the mean terrace width. This further indicates that adjacent steps frequently approach each other.
Hence, the deviation from the equilibrium TWH occurs for $\ell_0^{(111)}< \sigma$.
However, it is difficult to see in Fig.~\ref{fig_terraceW} (a) around $\sigma \equiv \ell_0^{(111)}$, because the deviation is small.
This distinction between the onset condition and its observable manifestation is essential for correctly interpreting the TWH.

%characteristic of the (001) terrace W

For the (001) terrace width, the histogram shape changes in a complex manner.
Eq. (\ref{eq_histo}) no longer approximately describes the TWH.
At $\Delta \mu/\epsilon = 0.6$, the frequencies of the TWH around $40<\ell/a$ increase from their equilibrium values.
The surface at $\Delta \mu/\epsilon = 0.6$ is situated at the boundary between the BKT and KPZ-like1 regions in the kinetic roughening diagram, Fig.~\ref{kr-diagram} (b).
Then, at $\Delta \mu/\epsilon = 0.8$, the TWH overlaps the equilibrium histogram
(BKT rough region).
At $\Delta \mu/\epsilon = 1.4$, where the surface is located in the KPZ-like1 kinetic rough region in Fig.~\ref{kr-diagram} (b), the frequencies of the TWH around $40<\ell/a $ increase significantly compared to the equilibrium values and also overlap with the data at $\Delta \mu/\epsilon = 0.6$.
Assuming a typical lattice constant of $a = 0.4$ nm, $40a$ corresponds to $16$ nm on the (001) terrace.
At the same time, the frequencies of $\ell <\ell_0$ decrease slightly.
When the surface is KPZ-like1 rough, the frequencies of the wide terraces decrease as $\ln N(\ell) \sim - c'_2 \ \ell$ as $\ell$ increases,
where $c'_2$ is a fitting constant for the (001) terrace-width tail.

%dmu =2.6 at (001)
At $\Delta \mu/\epsilon = 2.6$, where $\Delta \mu$ is larger than $\Delta \mu_{c\rm{MC}}^{(001)}$, (001) steps are not defined, because the (001) surface is atomically and thermodynamically rough.
However, the (111) steps remain well defined because the (111) terrace is always atomically and thermodynamically smooth.
KPZ-like3 rough surfaces exist at high temperatures and large $\Delta \mu$ (Figs.~\ref{kr-diagram} (a) and (b)).

%conclusion of this sec.
As demonstrated by the TWHs, terrace width inhomogeneity emerges spontaneously
in the KPZ-like1 and KPZ-like2 regions in the kinetic roughening diagrams.

\subsection*{Intrinsic step jamming}

Intrinsic step jamming refers to the spontaneous emergence of step density inhomogeneity under nonequilibrium driving, resulting from the interplay between asymmetric atomic processes and the non-penetrability constraint imposed by the RSOS model.
This inhomogeneity is directly reflected in the terrace width histogram (TWH), providing an observable signature of the phenomenon.
%specific quantities
The characteristic features of the simulation data are 
1) the roughness exponent $\alpha$ equals the KPZ value $\alpha_{\rm KPZ}= 0.3869$; 
2) SHD $P(\Delta h)$ has non-negligible skewness associated with surface step density; 
3) the sign of the skewness corresponds to bell- or cup-shaped surface undulations; 
4) TWHs exhibit exponential tails for large terrace widths, such that
$\ln N(\ell) \sim - c'_2 \ell$ for (001) terraces and
$\ln N(\ell^{(111)}) \sim - c_2 \ell^{(111)}$ for (111) terraces;
for the latter, see Eq. (\ref{eq_histo});
5) the TWH does not exhibit the form expected from long-range step--step repulsion.

\subsubsection*{ASEP-like step jamming in KPZ-like2 rough stepped surfaces}

This subsubsection explains how intrinsic step jamming occurs and how it creates cup-shaped surface undulations (Fig.~\ref{surfdatCut}) by forming a wide terrace in the stepped surface.
Here, the upper panels of Fig.~\ref{surfdatCut} (a) and (b) show an overhead view of the surface, while the lower panels display the side view along the bottom line in the upper panel.
In the overhead view, surface height is shown by brightness, with 10 levels, with higher brightness corresponding to higher surface height.
When the number of surface steps exceeds 10, the color of the surface layer next to the darkest one becomes the brightest. 
In the side view, the light green lines indicate coarse-grained cup- or bell-shaped structures.

%KPZ-like2 = ASEP
In the KPZ-like2 regime, the surface evolution is governed by quasi-1D step flow on large scales, where steps propagate predominantly in a common direction while remaining, on average, straight.
However, from a microscopic viewpoint,
transient local inhomogeneity arises in the step density.
This geometry allows the nanoscale step dynamics to be interpreted as an ASEP-like congestion process, with step impenetrability imposed by the SOS restriction.

%about ASEP
The ASEP is a paradigmatic model of a one-dimensional driven system with biased diffusion of hard-core particles  \cite{blythe}. 
The hard-core particles cannot penetrate each other and advance with probability $p$ and retreat with probability $q \neq p$.
In the case of an open system, the ASEP has a long history and first appeared in the literature as a model of biopolymerization and membrane transport.
Over the years, applications to other transport processes have emerged, e.g., as general models for traffic flow and in various theoretical and experimental studies of biophysical transport.
%Frank and Asep
Early work by Frank \cite{frank58} showed, using a 1D continuous model, that step bunching can arise as kinematic waves associated with step conservation and a slope-dependent surface kinetic coefficient, leading to shock-like accumulation.
This scenario closely resembles the continuous theory of the 1D ASEP \cite{blythe}, where high- and low-density states of particles spontaneously form in front and at the back of the shock.

%multi-lane
The KPZ-like2 rough surface is not exactly the 1D ASEP model, but analogous to a quasi-1D one \cite{ishiguro}.
If a point on the step edge is regarded as a hard-core particle,
the motion of a step edge corresponds to the motion of a particle.
The resulting congestion does not correspond to
the global step bunching envisioned in Frank's 1D model,
but rather to local step jamming.
The adjacent step edges correspond to the cars in the adjacent lanes of traffic, where cars are connected by chemical bonds.
Then, the neighboring parts of the step edge follow rapidly enough to form a connected jammed region.
Hence, the intrinsically step-jammed area in the running direction ($\langle -110\rangle$) of steps is on the nanoscale as shown in Fig.~\ref{surfdatCut}(a).

%(111) terrace; key factors
As shown in Fig.~\ref{fig_terraceW} (a), the frequencies of the wide (111) terraces increase for KPZ-like2 kinetic rough surfaces with $p=0.919$ as $\Delta \mu$ increases.
There are two key factors: one is asymmetric fluctuations during atom attachment and detachment at step edges; the other is the RSOS restriction (i.e., the SOS constraint with restricted height differences), where the SOS restriction indicates no overhangs.

%SOS restriction
When a wide terrace forms as shown in Fig.~\ref{surfdatCut} (c), assuming that the height of the left side is higher, the step on the higher side (colored in a thick blue) can advance to the right (grow); whereas the step on the lower side (colored in a thick red) cannot advance because its movement is blocked by the adjacent step beneath the red one due to the SOS restriction.
This restriction results in non-penetrability, or equivalently, an exclusion effect between steps.
For the retreat, the lower step can move back; however, the higher step cannot move back to the left due to the SOS restriction.

% asymmetry or symmetry fluctuations
When the crystal grows/recedes, the transition probabilities $w_{\rm trans}$ for the events shown in Fig.~\ref{surfdatCut} (d) and (e), such as the attachment and detachment of an atom, are given by $w_{\rm conf}\cdot w_{\rm drive}$, where $w_{\rm conf}$ and $w_{\rm drive}$ are $\exp[-\Delta E_{\rm conf}/\kBT]$ and $\exp[-\Delta E_{\rm drive}/\kBT]$,  respectively.
The energy changes are calculated from ${\cal H}_{\rm conf}$ and  ${\cal H}_{\rm drive}$ in Eq. (\ref{hamil}), respectively.
Since we adopt the Metropolis algorithm, $w_{\rm conf}\cdot w_{\rm drive} = 1$,  if $\Delta E_{\rm conf} + \Delta E_{\rm drive} \leq 0$.
We say that fluctuations caused by the probability $w_{\rm conf}$ are symmetric, while those caused by $w_{\rm drive}$ are asymmetric.
The asymmetric fluctuations cause the crystal to grow or retreat overall.

%cup-shaped, growth
Bell- or cup-shaped surface undulations are formed by intrinsic step jamming, as illustrated in Fig.~\ref{surfdatCut} (f) and (g), where arrows indicate the ``current'' $J$  direction, and the crosses indicate the prohibition of the motion by the RSOS restriction.
In crystal growth (Fig.~\ref{surfdatCut}(f)), when a wide terrace forms accidentally, the blue step advances faster than steps in layers above it.
Since the step density in layers above the blue step is higher than around the blue step, step advancement is frequently inhibited, resulting in a lower step that grows faster than steps in higher layers.
Therefore, cup-shaped surface undulations form around a wide terrace.

%bell-shaped, retreat
In the crystal step retreat (Fig.~\ref{surfdatCut} (g)), the red steps in Fig.~\ref{surfdatCut} (c) recede faster than lower-layer steps. 
This occurs because the step density in the lower layer is higher than that around red steps, frequently inhibiting step retreat.
Therefore, bell-shaped surface undulations form around the wide terrace. 

The bell- or cup-shaped surface undulations, in this way, lead to positive or negative skewness in the SHD.

%conclusion of this subsubsection
It should be emphasized that this intrinsic step jamming does not correspond to conventional step bunching extending over the entire system, but rather to transient traffic-jam-like clusters that continuously form and dissolve due to stochastic step dynamics.
Consequently, the jammed regions remain localized at the nanoscale, a key characteristic of intrinsic step jamming for KPZ-like2 rough surfaces.

%%%%% FIGURE  %%%%%

\begin{figure}%[h]%[htbp]
%\begin{center}
\centering
\includegraphics[width=8cm,clip]{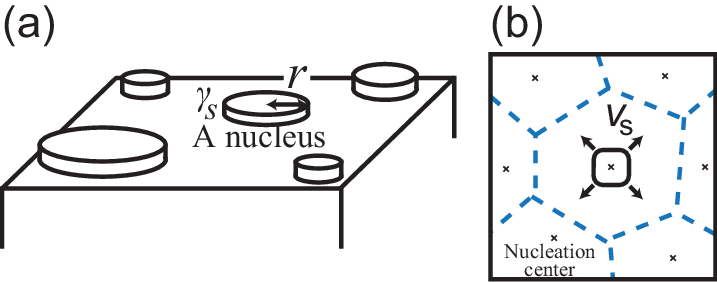}%
%\end{center}
\caption{Illustration of the poly-nuclear growth process on the (001) surface.
(a) A perspective view. 
(b) An overhead view.
}  
\label{fig_JMAK}
\end{figure}
%%%%% FIGURE  %%%%%

\subsubsection*{Intrinsic step jamming coupled with 2D poly-nuclear growth in KPZ-like1 stepped surfaces \label{sec_JMAK}}

While intrinsic step jamming in the KPZ-like2 regime is governed by quasi-1D ASEP-like step-flow dynamics, intrinsic step jamming in the KPZ-like1 regime emerges through coupling with two-dimensional (2D) poly-nuclear growth processes, leading to circularly stepped surface structures.
Thus, the underlying mechanism differs, but the resulting step density inhomogeneity is common to both regimes.
We next explain how intrinsic step jamming emerges in stepped surfaces and produces
bell-shaped undulations during crystal growth and cup-shaped undulations during crystal retreat
(Fig.~\ref{surfdatCut}(b), (h), (i)).
The cup- or bell-shaped surface undulations result in positive and negative skewness in the SHD, respectively, leading to KPZ-like1 kinetic roughness.

%2D single nucleation
On the (001) surface and its vicinal surfaces,
adatoms, vacancies, and their clusters form on terraces.
For crystal growth, the 2D mono-nuclear
growth process is generally regarded as the growth
mechanism on atomically smooth surfaces \cite{bcf}.
If we assume the shape of the adatom island to be a circle with radius $r$,   the free energy of an island is expressed as $G= -\pi r^2 \Delta \mu/a^2 + 2\pi r \gamma_s$,  where $\gamma_s$ is the step free energy per length or the step tension (Fig.~\ref{fig_JMAK}).
The maximum value of $G$, denoted by $G^*$, is obtained as
$G^* = \pi \gamma_s^2 a^2/\Delta \mu$ at the critical nucleus
size $r^* = \gamma_s a^2/\Delta \mu$.
The surface growth rate in a single nucleation process $V_1 \propto a I_n(\Delta \mu)$ with nucleation rate $I_n(\Delta \mu) \propto \exp [-G^*/\kBT] = \exp [-g^*/\Delta \mu]$.
%JMAK
The surface growth rate for a 2D poly-nuclear growth process $V_{\rm poly}$ is modeled according to the Johnson--Mehl--Avrami--Kolmogorov (JMAK) theory \cite{saito, ookawa, ramos, rikvold, novotny}:
\beqa
V_{\rm poly}  \approx && a \ v_{s}^{D/(D+1)} \  I_n(\Delta \mu)^{1/(D+1)} %\nonumber \\
%&& 
\propto \exp[-g'/\Delta \mu ],  \nonumber \\
 && g' = g^{*}/(D+1) , \quad  D=2,  \label{eq_v_poly}
\eeqa
where $v_s$ is the step growth rate indicated in Fig.~\ref{fig_JMAK} (b).

%Breakdown of JMAK theory
In our previous works \cite{akutsu25-3}, we found that our data are well explained by a mono-nuclear growth process, with step tensions calculated by the 2D Ising model near equilibrium; in the poly-nuclear growth process region, our data are overall consistent with the JMAK theory, but the theory breaks down in certain details.
The breakdowns are as follows: 1) the crossover point from the 2D mono-nuclear growth region to the 2D poly-nuclear growth region converges to the KPZ-like1 kinetic roughening transition point in the limit of $L \rightarrow \infty$  within numerical error for every temperature; 2) in the 2D poly-nuclear growth region, the data show that $g'$ is approximately $g^{*}/2$  in this region, rather than $g^{*}/3 $ \cite{akutsu24-2}.
We demonstrated that introducing the restricted geometry effect for the bell-shaped surface undulations caused by the mixing of island-on-island structures (Fig.~\ref{surfdatCut} (b), (h), (i)) can explain the increase in $g'$ \cite{akutsu25-3}.

%%%%% FIGURE  %%%%%

\begin{figure}%[h]%[htbp]
%\begin{center}
\centering
\includegraphics[width=7cm,clip]{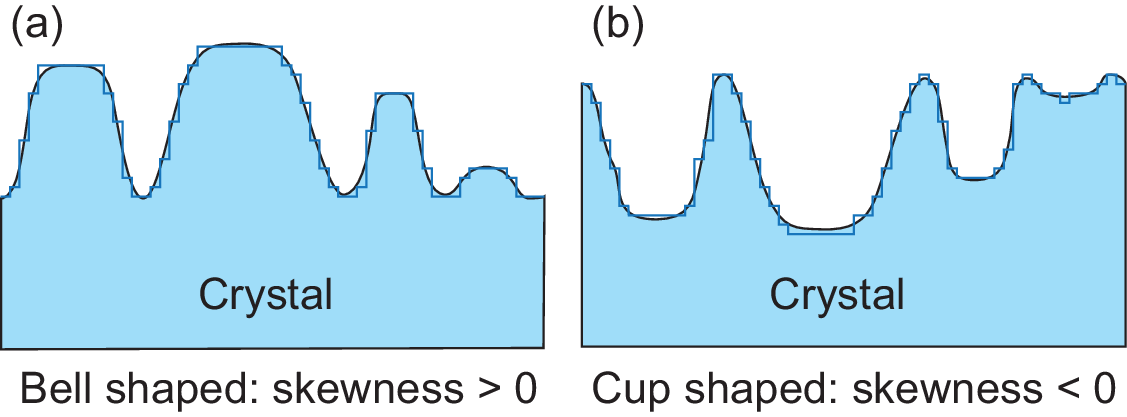}%
%\end{center}
\caption{
Schematic side views of surface undulations generated by intrinsic step jamming.
(a) Bell-shaped surface undulation,
leading to positive skewness in the surface-height-difference distribution (SHD).
(b) Cup-shaped surface undulation,
leading to negative skewness in the SHD.
}  
\label{fig_bell_cup_shaped}
\end{figure}
%%%%% FIGURE  %%%%%

%%%%% FIGURE  %%%%%

\begin{figure*}%[h]%[htbp]
%\begin{center}
\centering
\includegraphics[width=16cm,clip]{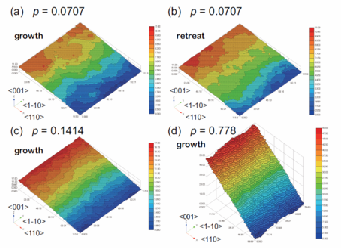}%
%\end{center}
\caption{Perspective views of the inclined surfaces.
$\kBT/\epsilon = 0.4$.  $4 \times 10^8$ MCS/site.
(a) KPZ-like1 rough, $\Delta \mu/ \epsilon = 1.4$, $L=80 \sqrt{2}a$.
(b) KPZ-like1 rough.  $\Delta \mu/ \epsilon = -1.4$, $L=80 \sqrt{2}a$.
(c) BKT rough, $\Delta \mu/ \epsilon = 1.4$, $L=80 \sqrt{2}a$. 
(d) KPZ-like2 rough, $\Delta \mu/ \epsilon = 1.4$, $L=40 \sqrt{2}a$. 
}  
\label{fig_DD3surf}
\end{figure*}
%%%%% FIGURE  %%%%%

%for referee comments
%nucleation as the selection of fluctuations

In crystal growth, thermal fluctuations generate adatoms and vacancies.
Because of the asymmetric transition probability $w_{\rm drive}$,
atomic attachment occurs more frequently than detachment.
Atomic attachment and detachment events occur
through the transition probabilities $w_{\rm conf}$ and  $w_{\rm drive}$.
After many events, adatom clusters that are nearly circular on average
are statistically selected by $w_{\rm drive}$ and tend to increase their area
(Figs.~\ref{surfdatCut}(b),(d)).
The peripheral energy of the hole cluster is the same as that of the adatom cluster with the same shape.
This leads to the same $w_{\rm conf}$.
However, $w_{\rm drive}$ selects not the hole cluster but the adatom cluster to increase the crystal area.
Individual atomic events are stochastic,
yet their statistical outcome is well described
by the 2D nucleation theory.

%island-on-island structures
When $\Delta \mu$ becomes large, steady growth in the RSOS model
can form island-on-island structures (Fig.~\ref{surfdatCut}(h)).
This contrasts with the 2D Ising model \cite{ramos, rikvold, novotny},
where nucleation and growth relax from a metastable state
to a stable state.
The average distance between adjacent circular steps is determined by the statistical survival rate of a cluster, which is described by the 2D nucleation rate $I_n(\Delta \mu)$.
At low temperatures, statistical selection caused by $w_{\rm conf}$ is relatively strong due to
the large free energy of steps, resulting in a relatively large average distance between adjacent steps.
The distance decreases as $\Delta \mu$ or temperature increases.

%->step jamming
When the islands that form on the terraces between steps grow, one side of the island is absorbed by the step at the same height
to form a wide terrace,
whereas the opposite side is blocked by the lower step
due to the SOS restriction, forming step congestion.
In this way, intrinsic step jamming also occurs around island-on-island structures,
thereby locally reducing the effective geometric dimension
of the surface from two to one \cite{akutsu24-2}.
This blending of step-jammed structures on the surface increases $g'$ \cite{akutsu25-3}.

%step jamming
Intrinsic step jamming causes bell-shaped surface undulations during crystal growth (Figs.~\ref{surfdatCut}(b), \ref{fig_bell_cup_shaped}(a)).
When the step growth rate $v_s$ is large, and the step distance between adjacent steps is large, the step growth rate of top-layer steps becomes larger
than that of lower-layer steps
because the number of blocking points increases
in the lower layers.
Due to the $w_{\rm drive}$ effect, the crystal tends to maximize its volume (Fig.~\ref{surfdatCut}(d) and (e)).
Hence, bell-shaped surface undulations occur (Fig.~\ref{surfdatCut}(b),(h)).   
The bottom layer is shaped like a concave corner with circles cut out (Fig.~\ref{surfdatCut} (b)).

%surface retreat
In the case of the surface retreat on the (001) and its vicinal surfaces, atomic vacancies occur more frequently than atomic adsorption by thermal fluctuation due to $w_{\rm drive}$ (Fig.~\ref{surfdatCut} (d), (e)).
As a result of statistical selection, a 2D island of holes (a hole island) forms more frequently than a 2D island of adatoms.
When $\Delta \mu$ is large, the hole-in-hole structure is created (Fig.~\ref{surfdatCut} (i), Fig.~\ref{fig_bell_cup_shaped}(b), Fig.~\ref{fig_DD3surf} (b)) for the steady crystal retreat.
As the hole islands formed on the terraces between steps grow, the lower step side of the hole islands is absorbed to form a wider terrace in a layer below; 
whereas the upper step side of the hole islands is blocked by the step due to the SOS restriction (Fig.~\ref{surfdatCut} (i)).
Since the number of block points of the step on the higher layer is larger than that of the step on the lower layer, the step retreat rate of the upper layer is slower than that of the lower layer.
Due to the $w_{\rm drive}$ effect, the crystal tends to minimize its volume (Fig.~\ref{surfdatCut}(d) and (e)).
Thus, the hole-in-hole structure has a cup-shaped form (Fig.~\ref{fig_DD3surf} (b)), resulting in negative skewness in SHD.

%large delta mu
 When the 2D nucleation rate $I_n(\Delta \mu)$ is large enough to be comparable to the step growth (retreat) rate, the bell- (cup-) shaped surface undulations formed by steps do not occur for crystal growth (retreat). 
 In that case, surface steps are not well discernible.
Hence, for $\Delta \mu_{c,{\rm MC}}^{(001)}< \Delta \mu$, the KPZ-like1 or power-law kinetic rough surface crosses over to a kinetic rough surface with adhesive growth.

%!!
%(001) surface with symmetric fluctuations
At high temperatures near equilibrium, adatom and hole clusters are frequently formed by symmetric thermal fluctuations with the transition probability $w_{\rm conf}$ and the clusters resist intrinsic step jamming.
The mean coherent length of the adatom or hole clusters formed by symmetric thermal fluctuations with the transition probability $w_{\rm conf}$ is approximately equal to the height-height correlation length $\xi$ at equilibrium, which is $0.6 a$, $a$, $2a$, and $4a$ for $\kBT/\epsilon =0.4$, 0.63, 0.83, and 1.1, respectively \cite{akutsu24-2} (Sec.~Methods).
The relatively small-sized holes (adatoms) prevent the step from advancing (or retreating) due to SOS restrictions.
Hence, the small clusters of adatoms or holes separate the locally congested steps.
The KPZ-like1 region shrinks to disappear as temperature increases in the kinetic roughening diagram (Fig.~\ref{kr-diagram} (a)).

\subsubsection*{Remarks}

This subsubsection clarifies the applicability, limitations, and experimental relevance of the intrinsic step jamming mechanism identified in this study.

%model's reliability
The present study focuses on isolating the mechanism of intrinsic step jamming under interface-limited conditions using the RSOS model with asymmetric attachment and detachment kinetics.
Within this framework, the numerical results are obtained
with high statistical accuracy and provide a reliable
description of the microscopic mechanism that generates
step density inhomogeneity on KPZ-like kinetically
roughened crystal surfaces.

%experimental possibilities
%X-ray
To observe the roughness exponent in the steady state, X-ray \cite{sinha} or light scattering \cite{harada89} in the crystal truncation rod (CTR) method \cite{robinson} is applicable \cite{akutsu25-2}.
In that situation, the kinetic roughening diagrams suggest that the temperature $T$, driving force for crystal growth $\Delta\mu$, and surface slope $p$ should be fixed; otherwise, the measured roughness exponent becomes an averaged value over surfaces with different roughness exponents.
%SHD TWH
It is possible to determine the KPZ-like roughness by sampling using an atomic force microscope (AFM) or a scanning tunneling microscope (STM), constructing a TWH, and measuring the skewness of the SHD.

%Complex real system
%
Real crystal surfaces generally involve additional
processes such as surface or volume diffusion, elastic interactions,
and environmental effects.
%epitaxial growth, vapor phase
In the case of epitaxial growth from the vapor phase, where surface diffusion is crucial, the universality class of the surface roughness changes.
The surface roughness is known to obey the Villain--Lai--Das Sarma (VLDS) class \cite{villain91, wolf91, petrov08, reis10, petrov15}, where the exponents are $\alpha=(4-d)/3$, $\beta= (4-d)/(8+d)$, and $z=(8+d)/3$.
However, in crystal growth by the sublimation method, where surface diffusion can be sufficiently fast compared to attachment-detachment kinetics, so that mass transport on terraces is not rate-limiting, the overall growth dynamics is controlled by step-edge processes.
In that case, surface kinetics are significant alongside surface diffusion, and intrinsic step jamming may affect the surface kinetic coefficient.

%melt growth

Intrinsic step jamming refers to the step density inhomogeneity that remains even after excluding the origins that cause conventional step bunching.
In the case of melt or solution growth, the KPZ-like rough surface caused by the intrinsic step jamming can be crucial for morphologies such as crystal growth shape (CGS) through the anisotropic surface kinetic coefficient. 
In our previous work \cite{akutsu25}, the calculated mesoscopic CGS successfully described the qualitative features of reversible shape changes in Si crystallites during growth from and retreat into the Si liquid.
The surface evolution is calculated based on the M\"{u}ller-Krumbhaar, Burkhardt, and Kroll's equation \cite{burkhardt}, which was derived under the consideration of the time-dependent Ginzburg--Landau theory, using the slope-dependent kinetic coefficient and surface stiffness of this RSOS model on the tangential plane at the nanometer scale.

\subsection*{Suppression of intrinsic step jamming}

Intrinsic step jamming occurs due to the presence of $w_{\rm drive}$, which induces nonequilibrium driving in the system.
Therefore, it is difficult to eliminate the intrinsic step jamming.
Fortunately, three distinct strategies can be identified to suppress intrinsic step jamming in kinetic roughening.

%%by way of the specific surface gradient
One way to suppress intrinsic step jamming is to find a specific surface gradient $p = \tan \theta$, where $\theta$ is the off-angle from the (001) surface tilted toward the (111) surface, that exists in the BKT rough region between the KPZ-like1 and the KPZ-like2 regions.
The skewness of the SHD is positive in the KPZ-like1 region  (Fig.~\ref{fig_DD3surf} (a), (b)), while it is negative in the KPZ-like2 region  (Fig.~\ref{fig_DD3surf} (d)).
There is a narrow window where the skewness of the SHD is near zero; for example, $p=0.1414$ in Fig.~\ref{kr-diagram} (b).
The surface with $p=0.1414$ is shown in Fig.~\ref{fig_DD3surf} (c).
$gW^2/\ln L$ is approximately 0.05, which is about half the value at $\TR^{001}$ (Fig.~\ref{kr-diagram} (g)). 
On this slope, the cup-shaped and bell-shaped surface undulations cancel each other out, thereby suppressing intrinsic step jamming.
The surface will be suitable for step-flow growth.

%how to determine the specific surface gradient
From an experimental viewpoint, the specific surface slope associated with the BKT rough regime can, in principle, be identified in situ by monitoring surface roughness over a wide area using crystal truncation rod (CTR) scattering \cite{akutsu25-2, sinha, robinson, harada89}.
Based on the slope dependence of $V$, the specific surface gradient that exists in the BKT rough region will be found as the inflection point of the surface growth rate $V$ with respect to $p$, that is, $\partial ^2 V/\partial p^2 = 0$.
Referring to the $p$ dependence of $V$, which was calculated in Refs.~\cite{akutsu23, akutsu25} for several temperatures, we find that bell-shaped undulations occur when $\partial^2 V/\partial p^2$ is positive, whereas cup-shaped undulations occur when $\partial^2 V/\partial p^2$ is negative.
This is consistent with the results obtained by Frank \cite{frank58} using a continuous model.
However, it is not very accurate for determining the inflection point on the $V$-$p$ curve \cite{akutsu23}. 
It is more accurate for determining the specific surface gradient from the asymptotic behavior of $W^2$ as shown in Fig.~\ref{kr-diagram} (g) and (h).
Using X-ray \cite{sinha, robinson} or light scattering \cite{harada89} under the crystal truncation rod (CTR) condition \cite{robinson, harada89},  the specific surface slope can be found \cite{akutsu25-2}.

%temperature increase
Another way to suppress intrinsic step jamming is to increase the temperature.
The relatively small islands of adatoms created by symmetric thermal fluctuations with $w_{\rm conf}$ hinder step retreat, while hole islands block step advancement.
In this way, the small clusters of adatoms or holes formed by thermal fluctuations with $w_{\rm conf}$ disperse the step positions.

%decrease driving force 
The remaining way to suppress intrinsic step jamming is to decrease the driving force $|\Delta \mu| <  \Delta \mu_{cr} = 0.3 \epsilon$ at $\kBT/\epsilon =0.4$ \cite{akutsu23}.
This method is the simplest; however, productivity drops.
The surface growth rate $V$ converges to 0.0206,  0.0261, and 0.0395 $a$/(MCS/site) for $p= 0$, 0.07070, 0.1414, respectively, at $\Delta \mu /\epsilon =1.4$.
While for $\Delta \mu /\epsilon =0.2$, $V$ becomes 0, 0.00640, 0.0129, 0.0192, and 0.0366 $a$/(MCS/site) at $p= 0$, 0.0884, 0.177, 0.265, and 0.7071, respectively.

%\subsection{Future outlook}

%actual crystal growth
%In this subsubsection, 

\section*{Methods}
\subsection*{Terrace width for (111) steps}

The terrace width for a regular train of (111) steps $\ell^{(111)}$ is calculated from the step height $n$ (Fig.~\ref{fig_deltahDef}).
The unit lengths of a unit cell are $a/\sqrt{2}$ in the $\langle 110 \rangle$ direction and  $a$ in the $\langle 001 \rangle$ direction.
Then, the length of the diagonal equals $\sqrt{3/2}\, a$.
$\ell^{(111)}$ for the regular train of steps with terrace size $n$ equals $\ell^{(111)} = \sqrt{3/2}\, n a$.
On the other hand, the surface gradient $p$ of the regular train of steps with terrace size $n$ is expressed as $p = \sqrt{2} \, n/(1+n)$,  where 1 on the right-hand side is for the (001) side surface of the (111) step.

\subsection*{Estimation of the correlation length at equilibrium}

It is exactly established that the inverse spin-spin correlation length for the 2D nearest-neighbor (nn) Ising model equals the interface tension (the interface free energy per length) at the boundary of the spin-up and spin-down domains \cite{baxter}.
At sufficiently low temperatures, the structure of a surface step is equivalent to the structure of the domain boundary in 2D nn Ising model.
The interface tension of the 2D nn square Ising model is calculated by the imaginary path-weight random walk (IPW) method \cite{akutsu92, akutsu90prl, akutsu99, akutsu14-2}, which is an extended method of Vdovichenko's method \cite{vdovichenko} to obtain the exact free energy of the Ising model.
Up to $\kBT/\epsilon = 0.83$, the correlation length of the RSOS model is obtained as the inverse of the interface tension of the 2D nn Ising model calculated by the IPW method.

Since the transition temperature $\kBT_c/\epsilon = 1/\ln[1+\sqrt{2}] \approx 1.13$, the correlation length at $\kBT/\epsilon =1.1$ is obtained as the inverse of the step tension of the RSOS model at equilibrium \cite{akutsu98, akutsu09, akutsu11JPCM} using the product wave-function renormalization group (PWFRG) method \cite{pwfrg, pwfrg2}, which is a transfer matrix version of the density-matrix renormalization group (DMRG) method \cite{dmrg} or the tensor network method \cite{tensorNetwork}.

\section*{Conclusion}

%\begin{itemize}

%(discovery  )
In this work, we investigated intrinsic step jamming on KPZ-like
kinetically roughened crystal surfaces using Monte Carlo simulations
of the RSOS model. 
Intrinsic step jamming occurs spontaneously at the nanometer scale.
As evident from the terrace width histogram, this phenomenon can occur even at terrace widths as short as 1.6\,nm.

%mechanism
We demonstrated that intrinsic step jamming originates from
asymmetric fluctuations in atomic attachment and detachment
under the SOS restriction. 
These asymmetric fluctuations
statistically select step configurations that generate
step density inhomogeneity. 
For KPZ-like1 (KPZ-like2) rough surfaces, the resulting inhomogeneity produces
bell- (cup-) shaped surface undulations during crystal growth and
cup- (bell-) shaped undulations during crystal retreat, leading to
positive or negative skewness in the surface-height-difference distribution or the local slope distribution.

%two regimes
Two distinct microscopic mechanisms of intrinsic step jamming were identified.
In the KPZ-like2 regime, step dynamics can be interpreted
as an ASEP-like congestion process arising from step impenetrability.
In contrast, in the KPZ-like1 regime, intrinsic step jamming
is coupled with two-dimensional poly-nuclear growth processes
associated with nucleation and statistical selection of
surface clusters.

%(key feature + implication)
The jammed regions do not develop into macroscopic step bunches extending across the entire system.
Instead, they remain localized on the nanoscale and
appear as transient clusters, analogous to traffic jams that repeatedly form and dissolve due to stochastic fluctuations.
This transient traffic-jam-like behavior represents a
key characteristic of intrinsic step jamming.

These results provide a statistical physics interpretation
of nanoscale step congestion on kinetically roughened
crystal surfaces. 
They also suggest that the emergence
of the BKT rough region in the kinetic roughening diagram
is not accidental but arises naturally from intrinsic
step jamming and the resulting step density inhomogeneity. 
Controlling the surface slope and step density may therefore provide a practical route to
stabilize desired surface morphologies during nanoscale crystal growth.

%\end{itemize}

\section*{Acknowledgements}

The author wishes to acknowledge Dr. Y. Ishiguro, Prof. Y. Suzuki, Prof. T. Sasada, and Assoc. Prof. K. A. Takeuchi for their valuable insights.
This work was supported in part by KAKENHI Grant-in-Aid (Nos. JP22K03487 (N.A.), JP24H00432 (Y.K.)) from the Japan Society for the Promotion of Science (JSPS)
and in part by JST SICORP Grant Number JPMJSC22C1 (Y.K.).
 The author acknowledges the use of ChatGPT (OpenAI) for assistance with English language editing and proofreading of the manuscript. All scientific content, interpretations, and conclusions are solely those of the author.

\section*{Author declarations}
\textbf{Conflict of Interest}\\
The authors declare no conflicts of interest.

\noindent
\textbf{Author contributions statement}\\
%Must include all authors, identified by initials, for example:
N.A. conceived and conducted the Monte Carlo calculations, analyzed the results, and wrote the main manuscript.
Y.K. provided expertise on epitaxial growth.

\section*{Data Availability}
%The data that support the findings of this study are publicly available in a GitHub repository at
%\url{https://github.com/noriko-akutsu/kpz-step jamming}.

%The data and simulation codes that support the findings of this study are publicly available in a GitHub repository at
%\url{https://github.com/noriko-akutsu/kpz-step jamming}.

The datasets used and/or analyzed during the current study are available from the corresponding author upon reasonable request.

%\section*{Supporting information}

%A listing of the contents of each file supplied as Supporting Information
%should be included. For instructions on what should be included in the
%Supporting Information as well as how to prepare this material for
%publications, refer to the journal's Instructions for Authors.

%The following files are available free of charge.
%\begin{itemize}
  %\item Filename-1: brief description
  %\item Filename-2: brief description
%\end{itemize}

%%%%%%%%%%%%%%%%%%%%%%%%%%%%%%%%%%%%%%%%%%%%%%%%%%%%%%%%%%%%%%%%%%%%%
%% If you are using classical BibTeX rather than biblatex,
%% remove the \printbibliography and uncomment the \bibliograpy one
%%%%%%%%%%%%%%%%%%%%%%%%%%%%%%%%%%%%%%%%%%%%%%%%%%%%%%%%%%%%%%%%%%%%%
%\printbibliography
%\bibliography one%{acs-template.bib}

\end{document}